\documentclass[twocolumn,showpacs,preprintnumbers,prb,superscriptaddress,floatfix]{revtex4}
\usepackage{epsfig,amssymb,amsmath,hyperref,bm}
\usepackage{graphics,graphicx,psfrag}

\newcommand\beq{\begin{equation}}
\newcommand\eeq{\end{equation}}
\newcommand\bea{\begin{eqnarray}}
\newcommand\eea{\end{eqnarray}}
\newcommand{\nonum}{\nonumber}

\newcommand\px{\partial_x}

\newcommand\tp{t_{\perp}}
\newcommand\tV{\tilde V}
\newcommand\tU{\tilde U}

\begin{document}

\title{From frustrated insulators to correlated anisotropic metals:
charge ordering and quantum criticality in coupled chain systems}

\author{Siddhartha Lal}
\email{slal@ictp.it}
\affiliation{The Abdus Salam ICTP. Strada Costiera 11, Trieste 34014, Italy.}  
\author{Mukul S. Laad}
\email{mukul@mpipks-dresden.mpg.de}
\affiliation{Max-Planck-Institut f\"ur Physik Komplexer Systeme, 01187 Dresden, Germany.}


\begin{abstract}
A recent study revealed the dynamics of the charge sector of a 
one-dimensional quarter-filled electronic system with extended Hubbard 
interactions to be that of an effective pseudospin transverse-field Ising 
model (TFIM) in the strong coupling limit.  
With the twin motivations of studying the co-existing charge and 
spin order found in strongly correlated chain systems and the 
effects of inter-chain couplings, we investigate the phase diagram 
of coupled effective (TFIM) 
systems. A bosonisation and RG analysis for a two-leg TFIM ladder 
yields a rich phase diagram showing Wigner/Peierls charge order
and Neel/dimer spin order. In a broad parameter 
regime, the orbital antiferromagnetic phase is found to be stable. 
An intermediate gapless phase of finite width is found to lie in 
between two charge-ordered gapped phases. Kosterlitz-Thouless 
transitions are found to lead from the gapless phase to either of the 
charge-ordered phases. A detailed analysis is also carried out for the 
dimensional crossover physics when many such pseudospin systems are 
coupled to one another. Importantly, the analysis reveals the key role 
of critical quantum fluctuations in driving the strong dispersion in 
the transverse directions, as well as a $T=0$ deconfinement transition. 
Our work is potentially relevant for a unified description of a class 
of strongly correlated, quarter-filled chain and ladder systems.
\end{abstract}



\maketitle
\section{Introduction}
Strongly correlated ladder systems are fascinating candidates 
for studying the interplay of spin and charge ordering, and their 
combined influence on the emergence of novel ground states~\cite{sachdev}. 
Several recent studies provide experimental 
realisations of such systems, exhibiting diverse phenomena
like charge order (CO), antiferromagnetism (AF) and unconventional 
superconductivity (uSC) as functions of suitable control 
parameters~\cite{tokura,yamauchi}.  
The existence of several non-perturbative theoretical techniques
in one dimension have also resulted in the study 
of the emergence of exotic phases from instabilities of the high-$T$ 
Luttinger liquid~\cite{giamarchi}.
Given that these systems are Mott insulators, longer-range 
Coulomb interactions are relevant in understanding CO/AF/uSC phases. 
Despite of some recent work~\cite{capponi},  
a detailed understanding of the effects of longer-range interactions in 
quasi-1D systems is a largely unexplored problem. 
Further, attention has mostly focused on studies of models at 1/2-filling
~\cite{giamarchi}. 
\par
Here, we study an effective pseudospin model describing charge degrees 
of freedom of a one-dimensional $1/4$-filled electronic system.  
Such an effective model can be 
derived from an extended Hubbard model (i.e., with longer-range interactions)
in a number of physically relevant cases. For example, it is reasonably 
well established that the structure of the material $Na_{2}V_{2}O_{5}$ 
is essentially that of a system of weakly coupled quarter-filled 
two-leg ladders ~\cite{fulde,mostovoy,horsch}. While the spin sector 
is effectively quasi-one dimensional with each ladder corresponding to 
a spin chain, the charge sector is well described by a one-dimensional 
system of Ising pseudospins in a transverse field. The pseudospin degrees 
of freedom correspond to the position of a localised electron on a given rung 
of the two-leg ladder. 
\par 
Another relevant example is the $Sr_{14}Cu_{24}O_{41}$ system, 
a ladder based material well described by a Hubbard-type fermionic model. Hitherto 
described by a Hubbard (or extended Hubbard) model at half-filling, 
recent experimental work
by Abbamonte {\it et al.}~\cite{sawatzky} strongly suggests a very new scenario.
Since pure $Sr_{14}Cu_{24}O_{41}$ is a Mott insulator
with short-ranged antiferromagnetic correlations and a spin gap, nearest 
neighbour (nn) Coulomb interactions are in fact a necessary ingredient of a 
minimal model.  Additionally, electronic charge-density-wave (e-CDW) order
is inferred for zero doping, providing additional evidence for inclusion of 
nn Coulomb interactions into a minimum effective model for these ladder 
systems. The e-CDW appears not to be driven by electron-phonon 
coupling, as shown by the absence of a detectable lattice distortion tied to the
e-CDW.  The relevance of inter-ladder coupling is also clearly 
revealed by this study.  
\par
Similar arguments hold for the quasi-one dimensional charge 
transfer organics, which show e-CDW order in the Mott insulating
phases in the generalised $T-P$ phase diagram (substitution of different
anions in the TMTSF-salts corresponds to varying chemical 
pressure)~\cite{jerome}. In studying a one-dimensional $1/4$-filled 
fermionic model with extended interactions, 
it has been suggested~\cite{emery,first} that the same Ising pseudospin 
Hamiltonian may be a relevant starting point for the study of charge-order 
in organics.  Clearly, a half-filled Hubbard (or extended Hubbard) model 
is inadequate when one seeks to understand Mott insulators with e-CDW (along 
with AF/dimerised) ground states.  A quarter-filled, extended Hubbard model 
turns out to be a minimal model capable of describing such ground states
~~\cite{yoshi1, yoshi2}. 
\par
While the weak coupling limit of the underlying extended Hubbard model
has been recently studied~\cite{donohue,orignac},
we note that the real systems under consideration are generically in 
the strong coupling regime of the model~\cite{fulde,mostovoy,horsch}. 
To the best of our knowledge, this regime has not been studied in sufficient 
detail. Another interest is to investigate the conditions under which 
short-coherence length superconductivity can arise by hole-doping a charge- 
and antiferromagnetically ordered Mott insulator. Again, this issue has 
been studied in sufficient detail only in the weak coupling limit~\cite{carr}.
Motivated by the above discussion, we have recently studied a 
problem of coupled electronic chains~\cite{first}, where each chain is 
described by an extended Hubbard model with a hopping term (of strength, $t$) 
and nearest-($V_{1}$) and next-nearest neighbor (nnn) ($V_{2}$) Coulomb interactions 
in addition to the local, Hubbard ($U$) interaction~\cite{first}. Further, 
we studied the strong coupling regime, where $U>>V_{1},V_{2}>>t$. In this 
regime, we derived an effective transverse-field Ising model (TFIM) in terms of 
pseudospins describing the charge degrees of freedom for a single chain
~\cite{first}. In this work, our primary aim will be to study the effects of 
interchain couplings in this system of chains. As noted earlier, it has also 
been shown in Ref.(\cite{mostovoy}) that 
a quarter-filled two-leg ladder in which the on-site 
Coulomb repulsion is the largest coupling can be mapped onto the same 
effective TFIM Hamiltonian. Our results will, therefore, also be relevant to 
an understanding of such ladder systems.
\par
We begin in section II, 
by briefly reviewing for the sake of completeness,
the derivation of the effective TFIM Hamiltonian for the charge sector of 
a system of $1/4$-filled  
electronic chains with strong extended Hubbard interactions and very weak 
nearest neighbour hopping. We refer readers to Ref.(\cite{first}) for a more 
detailed discussion. We then proceed 
with the effective pseudospin TFIM model for the charge 
degrees of freedom for a single chain~\cite{first} and couple 
neighboring chains to describe the physical situations detailed above. 
In this way, a system of two effective pseudospin models coupled to one 
another is first analysed 
in section II using abelian bosonisation and perturbative renormalisation 
group methods, to obtain a rich phase diagram showing different types of CO 
phases. At the same time, there remains open the intriguing 
possibility that some of these gapped phases may themselves be separated 
from one another by non-trivial gapless phases of finite width~\cite{tsvelik}. 
In what follows, we will encounter an example of this phenomenon in 
subsection IIB. The transitions from this gapless critical phase to either 
of the two charge-ordered phases are found to belong to the 
Kosterlitz-Thouless universality class. The  
question that naturally arises is whether such gapless, correlated, 
anistropic metallic phases survive when many such TFIM systems are 
coupled to one another so as to undergo a dimensional crossover. To 
answer this question reliably, we treat the inter-TFIM couplings  
at the level of the random phase approximation (RPA)~\cite{giamarchi,carr2,schulz} in 
section III. The finite $T$ neighbourhood of the quantum critical point (QCP) 
is studied in detail. 
Interestingly, the inter-TFIM hopping coupling is seen to drive 
the system away from a gapped phase at $T=0$ to a gapless one at  
th QCP: critical quantum fluctuations 
drive the system through a deconfinement transition together with a dimensional 
crossover.  
In section IV, we present a comparison of our findings with some 
recent numerical works. Finally, we conclude in section V.   
\section{Two-leg coupled TFIM ladder model}
We begin with the Hamiltonian for a spin chain system
\bea
H &=& -\sum_{n}[t(S_{n}^{x}S_{n+1}^{x} + S_{n}^{y}S_{n+1}^{y}) 
+ V S_{n}^{z}S_{n+1}^{z}\nonum\\ 
&& + P S_{n}^{z}S_{n+2}^{z} + hS_{n}^{z}]
\label{ham1}
\eea
where $S_{n}^{x}$, $S_{n}^{y}$ and $S_{n}^{z}$ and spin-1/2 operators. 
The couplings $(t, V,P) >0$ are the nearest neighbour (nn) XY, the 
nearest neighbour Ising and the next nearest neighbour (nnn) Ising couplings 
respectively and $h$ is the external magnetic field. For $h=0$, this 
Hamiltonian can be derived via a Jordan-Wigner transformation on the charge 
degrees of freedom of a $1/4$-filled extended 
Hubbard model of electrons on a one-dimensional lattice in the strong-coupling 
limit: the on-site Hubbard interaction $U\rightarrow\infty$, 
while the nn ($V$) and nnn ($P$) 
density-density interaction couplings and the nn electron hopping ($t$) 
are all taken to be finite~~\cite{first}
\beq
H = \sum_{n}[-\frac{t}{2}(c_{i}^{\dagger}c_{i+1}+h.c) 
+ V n_{i}n_{i+1} + P n_{i}n_{i+2}]~. 
\label{fermi}
\eeq
We study the problem in the 
limit of strong-coupling where $V, P >> t$ (but where 
$(V-2P)\sim 2t$)~~\cite{first}.
\par
Let us begin by studying the case of $t=0$~\cite{emery} (we 
will be studying eq.(\ref{ham1}) for the case of $h=0$ in all that 
follows). It is easy to see that for the case of $V>2P$, the 
ground state of the system is given by a Neel-ordered antiferromagnetic 
(AF) state with two degenerate ground-states given by
\bea
|AFGS1\rangle &=& |\ldots + - + - + - + - \ldots\rangle\nonum\\
|AFGS2\rangle &=& |\ldots - + - + - + - + \ldots\rangle
\label{wig1}
\eea
where we signify $S_{n}^{z}=1/2,-1/2$ by $+$ and $-$ respectively and we 
have explicitly shown the spin configuration in the site nos. 
$-3\leq n \leq 4$ in the ground states. In the original electronic 
Hamiltonian eq.(\ref{fermi}), this AF order corresponds to a Wigner 
charge-ordering (CO) 
in the ground-state. Similarly, for the case of $V<2P$, the 
ground state of the system is given by a dimer-ordered  
$(2,2)$ state \cite{emery} with four degenerate ground-states given by
\bea
|22GS1\rangle &=& |\ldots - + + - - + + - \ldots\rangle\nonum\\
|22GS2\rangle &=& |\ldots - - + + - - + + \ldots\rangle\nonum\\
|22GS3\rangle &=& |\ldots + - - + + - - + \ldots\rangle\nonum\\
|22GS4\rangle &=& |\ldots + + - - + + - - \ldots\rangle\nonum\\
\label{pei1}
\eea
where we signify $S_{n}^{z}=1/2,-1/2$ by $+$ and $-$ respectively and we 
have explicitly shown the spin configuration in the site nos. 
$-2\leq n \leq 5$ in the ground states. In the original electronic 
Hamiltonian eq.(\ref{fermi}), this $(2,2)$ order corresponds to a 
Peierls CO in the ground-state. 
\par
The effect of the XY terms in the Hamiltonian (\ref{ham1}) on these ground 
states is now considered. Let us start with noting the effect of a XY term 
on a single nn spin-pair on the 4 degenerate $(2,2)$ ground-states; for 
purposes of brevity, we will denote the entire 
$t (S_{n}^{x}S_{n+1}^{x} + S{n}^{y}S_{n+1}^{y})$ 
term simply as $t^{n,n+1}$. Thus, 
\bea
t^{0,1}|22GS1\rangle &=& t^{0,1}|\ldots + - \ldots\rangle 
= \frac{t}{2}|\ldots - + \ldots\rangle\nonum\\
t^{0,1}|22GS2\rangle &=& t^{0,1}|\ldots + + \ldots\rangle 
= 0\nonum\\
t^{0,1}|22GS3\rangle &=& t^{0,1}|\ldots - + \ldots\rangle 
= \frac{t}{2}|\ldots + - \ldots\rangle\nonum\\
t^{0,1}|22GS1\rangle &=& t^{0,1}|\ldots - - \ldots\rangle 
= 0\nonum\\
t^{-1,0}|22GS1\rangle &=& t^{-1,0}|\ldots + + \ldots\rangle 
= 0\nonum\\
t^{-1,0}|22GS2\rangle &=& t^{-1,0}|\ldots - + \ldots\rangle 
= \frac{t}{2}|\ldots + - \ldots\rangle\nonum\\
t^{-1,0}|22GS3\rangle &=& t^{-1,0}|\ldots - - \ldots\rangle 
= 0\nonum\\
t^{-1,0}|22GS4\rangle &=& t^{-1,0}|\ldots + - \ldots\rangle 
= \frac{t}{2}|\ldots - + \ldots\rangle~.
\label{flip1}
\eea
\par
In a similar manner, we study the action of the operator $t^{n,n+1}$ 
on the two degenerate ground states of the AF ordered configuration as
\bea
t^{0,1}|AFGS1\rangle &=& t^{0,1}|\ldots + - \ldots\rangle 
= \frac{t}{2}|\ldots - + \ldots\rangle\nonum\\
t^{0,1}|AFGS2\rangle &=& t^{0,1}|\ldots - + \ldots\rangle 
= \frac{t}{2}|\ldots + - \ldots\rangle\nonum\\ 
t^{-1,0}|AFGS1\rangle &=& t^{-1,0}|\ldots - + \ldots\rangle 
= \frac{t}{2}|\ldots + - \ldots\rangle\nonum\\
t^{-1,0}|AFGS2\rangle &=& t^{-1,0}|\ldots + - \ldots\rangle 
\hspace*{-0.1cm}= \frac{t}{2}|\ldots - + \ldots\rangle~. 
\label{flip2}
\eea
\par
Defining bond-pseudospins $\tau_{i}^{z} = (S_{i}^{z} - S_{i-1}^{z})/2$, 
$\tau_{i}^{+} = S_{i}^{+}S_{i-1}^{-}$ and 
$\tau_{i}^{-} = S_{i}^{-}S_{i-1}^{+}$ (which can be rewritten in terms 
of bond-fermionic operators in the original electronic Hamiltonian 
eq.(\ref{fermi}) as
$\tau_{i}^{z} = (n_{i} - n_{i-1})/2$, $\tau_{i}^{+} = c_{i}^{\dagger}c_{i-1}$ 
and $\tau_{i}^{-} = c_{i}c_{i-1}^{\dagger}$ respectively), we can 
write the four degenerate ground states of the $(2,2)$ ordered 
configuration in terms of these bond pseudospins as
\bea
|22GS1\rangle &=& |\ldots~ 0~  -~  0~  +~  0 \ldots\rangle\nonum\\
|22GS2\rangle &=& |\ldots\hspace*{0.1cm}  +~  0~  -~  0~  + 
\ldots\rangle\nonum\\
|22GS3\rangle &=& |\ldots~ 0~  +~  0~  -~  0 \ldots\rangle\nonum\\
|22GS4\rangle &=& |\ldots\hspace*{0.1cm}  -~  0~  +~  0~  - \ldots\rangle~,
\label{pei2}
\eea
where we have denoted $\tau_{n}^{z}=1/2$ as $+$ and $\tau_{n}^{z}=-1/2$ as 
$-$ and have explicitly shown the pseudospin configurations on the 
bond nos. $0\leq n \leq 4$. We can clearly see from eq.(\ref{pei2}) 
that these four 
ground-states break up into two pairs of doubly degenerate (AF) orderings 
of the pseudospins defined on the odd bonds ($|22GS1\rangle$ and 
$|22GS3\rangle$) on the even bonds ($|22GS2\rangle$ and 
$|22GS4\rangle$) respectively. It is also simple to see from eq.(\ref{flip1}) 
that the 
action of the operator $t^{n-1,n}$ (for the nearest neighbour pair 
of sites given by $(n-1,n)$) on these four ground states is to flip 
a pseudospin defined on the bond $n$ (lying in between the pair of 
sites $(n-1,n)$) or to have no effect at all. 
\par
We can now similarly see that the two-degenerate ground states of the AF 
ordered configuration can be written in terms of the bond-pseudospins 
defined above as
\bea
|AFGS1\rangle &=& |\ldots  +  -  +  -  + \ldots\rangle\nonum\\
|AFGS2\rangle &=& |\ldots  -  +  -  +  - \ldots\rangle
\label{wig2}
\eea
where we have explicitly shown the pseudospin configurations on the 
bond nos. $0\leq n \leq 4$. From eq.(\ref{wig2}), we see that the 
two degenerate 
ground-states have antiferromagnetic ordering of pseudospins on nn bonds; 
this can equally well be understood in terms of the ferromagnetic ordering 
of pseudospins on the odd bonds and on the even bonds separately. Further, 
from eq.(\ref{flip2}), we can see 
that the action of the operator $t^{n-1,n}$ (for the nearest 
neighbour pair of sites given by $(n-1,n)$) on these two ground states 
is again to flip a pseudospin defined on the odd (even) bond $n$ 
(lying in between the pair of sites $(n-1,n)$) against a background of 
ferromagnetically ordered configuration of pseudospins defined on the 
odd (even) bonds.
\par
Thus, we can model these pseudospin ordered ground states 
(\ref{pei2}),(\ref{wig2}) as well as {\it all} possible 
pseudospin-flip excitations above them (as given by action of  
operators of the type $t^{n-1,n}$ (\ref{flip1}),(\ref{flip2})) 
with the effective Hamiltonian~~\cite{first}
\bea
H &=& -\sum_{n\in odd}[2t\tau_{n}^{x} 
+ (V-2P)\tau_{n}^{z}\tau_{n+2}^{z}]\nonum\\
&& -\sum_{n\in even}[2t\tau_{n}^{x} 
+ (V-2P)\tau_{n}^{z}\tau_{n+2}^{z}]\nonum\\
&=& -\sum_{n}[2t\tau_{n}^{x} 
+ (V-2P)\tau_{n}^{z}\tau_{n+2}^{z}]~,
\label{ham2}
\eea
where $n$ is the bond index. 
\par
  This is just the Ising model in a transverse field, which is exactly 
solvable~\cite{lieb,niemeijer} and has been studied extensively in 1D
~\cite{sachdev,chakrabarti}.  If $(V-2P)>0$, the ground state is 
ferromagnetically ordered in $\tau^{z}$, i.e, it corresponds to a Wigner CDW.
 For $(V-2P)<0$, the Peierls dimer order results in the ground state.  At 
$(V-2P)<2t$, the quantum disordered phase
has short-ranged pseudospin correlations, and is a charge
``valence-bond" liquid.  
The quantum critical point at $(V-2P)=2t$ separating these phases is a
deconfined phase with gapless pseudospin ($\tau$) excitations, and 
power-law fall-off in 
the pseudospin-pseudospin correlation functions.  Correspondingly, 
the density-density 
correlation function has a power-law singular behavior at low energy, 
with an exponent
$\alpha=1/4$ characteristic of the $2D$ Ising model at criticality.
The gap in the pseudospin spectrum
on either side of the critical point is given by $\Delta_{\tau} = 2|V-2P-2t|$.  
Further, the quantum-critical behaviour extends to temperatures as
high as $T\sim \Delta_{\tau}/2$~~\cite{kopp} and undergoes finite-temperature
crossovers to the two gapped phases at $T\sim |\Delta_{\tau}|$.
The dynamics of the spin sector of a $1/4$-filled electronic system in the 
limit of strong correlations was also studied in Ref.(\cite{first}); we restrict 
ourselves to a few brief remarks on the spin sector in this work and direct 
the reader to Ref.(\cite{first}) for more details. Henceforth, we will focus primarily 
on the effects of interchain couplings on the charge sector of such chain systems.
\subsection{Bosonisation and RG analysis}
Thus, we now proceed with the effective pseudospin Hamiltonian for the charge sector
\beq
H^{chain} = -\sum_{j} [2t~\tau_{j}^{x} 
+   (V-2P)~\tau_{j}^{z}\tau_{j+1}^{z}]
\label{onechain}
\eeq
where the Ising pseudospin coupling $V-2P$ has 
been replaced by $V$ for convenience in all that follows.  
Rotating the pseudospin axis $\tau^{x}\rightarrow\tau^{z}$, 
$\tau^{z}\rightarrow - \tau^{x}$ and introducing a  
bond fermion repulsion 
$U_{\perp}\sum_{i,a,b\neq a}(n_{i,a}-n_{i+1,a})(n_{i,b}-n_{i+1,b})$ 
as well as a bond fermion transfer term 
$t_{\perp}\sum_{i,a,b\neq a}(c_{i,a,\uparrow}^{\dagger}c_{i+1,a,\uparrow}
c_{i,b,\downarrow}^{\dagger}c_{i+1,b,\downarrow} + {\rm h.c})$
between two such chain systems described by the indices ($a,b$)
(recall the relations between the pseudospins and the bond fermions in the 
original electronic model given earlier), 
we have the effective Hamiltonian for the charge sector of the 
coupled system in terms of a pseudospin ladder model
\bea
H 
&=& -\sum_{j,a}\lbrack 2t~\tau_{j,a}^{z} 
		+ V~\tau_{j,a}^{x}\tau_{j+1,a}^{x}\rbrack\nonum\\
&& -\hspace*{-0.2cm}\sum_{j,a,b\neq a} \lbrack 
U_{\perp}\tau_{j,a}^{z}\tau_{j,b}^{z} 
+ ~\tp~ (\tau_{j,a}^{x}\tau_{j,b}^{x} 
+ \tau_{j,a}^{y}\tau_{j,b}^{y})\rbrack~, 
\label{twochain}
\eea
where $a,b=1,2$ is the chain index. Having explored the strong coupling limit of 
$U_{\perp} >> V$ in an earlier work ~\cite{first}, we will explore 
the weak coupling scenario of $U_{\perp} << V$ below. 
We note that in the spin sector, coupling between the two systems leads to a $S=1/2$ 
Heisenberg ladder-type model, and has been studied extensively by 
several authors ~\cite{giamarchi,tsvelik}. 
Interchain spin coupling turns out to be relevant, opening up a spin 
gap: in this case two possibilities are known to result.  The ground state
has either short-ranged antiferromagnetic correlations (RVB) with no 
magnetic long-range order (LRO), or 
it spontaneously breaks translation symmetry, leading to dimerisation. 
\par
We now return to our analysis of the charge sector. To begin, we introduce 
fermionic operators $\psi_{j,a}$ on each chain via a Jordan-Wigner transformation 
of the pseudospins. We will then proceed to bosonise the theory. During this 
procedure, one has to take care that the spin commutation relations are 
maintained, i.e., the new fermionic operators have anticommutation relations 
on each chain but commute between chains. It can indeed be checked that 
the bosonised expression for the various pseudospin operators satisfy the 
required spin commutation relations~\cite{giamarchi,schulzmanychain}.
In terms of these fermions, upon denoting the chains as 
$a,b=\uparrow,\downarrow$, we find an effective Hamiltonian 
for the 1D Hubbard model with an equal-spin pairing term and an 
on-site spin-flip term
\bea
H &=& -\frac{{\tilde t}}{2}\sum_{j,a} (\psi_{j,a}^{\dagger}\psi_{j+1,a}
+ {\rm h.c})
-\frac{\tp}{2}\sum_{j}(\psi_{j,a}^{\dagger}\psi_{j,b} 
+ {\rm h.c})\nonum\\ 
&&\hspace*{-1.1cm} + \tU\sum_{j} n_{j,\uparrow} n_{j,\downarrow} 
+ \tV\sum_{j,a} (\psi_{j,a}^{\dagger}\psi_{j+1,a}^{\dagger} 
+ {\rm h.c}) + \mu\sum_{j,a} n_{j,a}
\label{effham}
\eea
where ${\tilde t}$ and $\tp$ are the in-chain and inter-chain 
hopping parameters respectively, $\mu=-2t$ the chemical 
potential of the effective model (and unrelated to the filling of 
the underlying electronic model), $\tU = - U_{\perp}$ the on-site 
(Hubbard) interaction 
coupling and $\tV = V/4$ the pairing strength. 
Note that while we treat the parameters ${\tilde t}$ and $\tV$ as 
independent parameters for the sake of generality, ${\tilde t}=\tV=V$ 
in our original model (\ref{twochain}).
Bosonising in the usual way (in terms of the 
usual charge $\rho = \uparrow + \downarrow$ and spin 
$\sigma = \uparrow - \downarrow $ variables), we obtain the 
effective low-energy bosonic Hamiltonian  
\bea
H &=& \frac{1}{2\pi}\int dx 
[v_{\rho} K_{\rho} (\pi \Pi_{\rho}(x))^{2} + \frac{v_{\rho}}{K_{\rho}} 
(\px \phi_{\rho} (x))^{2}]\nonum\\
&&\hspace*{-0.5cm} + \frac{1}{2\pi}\int dx
 [v_{\sigma} K_{\sigma} (\pi \Pi_{\sigma}(x))^{2} 
+ \frac{v_{\sigma}}{K_{\sigma}} (\px \phi_{\sigma} (x))^{2}]\nonum\\
&&\hspace*{-0.5cm} + \frac{U_{\sigma}}{2\pi\alpha}\int dx 
\cos(\sqrt{8}\phi_{\sigma}) 
+ \frac{U_{\rho}}{2\pi\alpha}\int dx \cos(\sqrt{8}\phi_{\rho})\nonum\\
&&\hspace*{-1.3cm} + \frac{\tV}{2\pi\alpha}\int dx 
\cos(\sqrt{2}\theta_{\rho})\cos(\sqrt{2}\theta_{\sigma}) 
- \frac{\sqrt{2}\mu}{\pi}\int dx \px\phi_{\rho} (x)\nonum\\
&&\hspace*{-1.3cm} + \frac{\tp}{\pi\alpha}\int dx \cos(\sqrt{2}\phi_{\sigma})
\cos(\sqrt{2}\theta_{\sigma}) + \frac{V_{1}}{\pi\alpha}\int dx 
\cos(\sqrt{8}\theta_{\sigma})\nonum\\
&&\hspace*{-1.3cm} + \frac{V_{2}}{\pi\alpha}\int dx \cos(\sqrt{2}\phi_{\sigma})
\cos(\sqrt{2}\theta_{\rho}) 
\eea  
where 
$\Pi_{\rho} = \frac{1}{\pi} \px\theta_{\rho}$~,~
$\Pi_{\sigma} = \frac{1}{\pi} \px\theta_{\sigma}$~,~
$v_{\rho} K_{\rho} = v_{F} = v_{\sigma} K_{\sigma}$~,~
$v_{\rho}/K_{\rho} = v_{F} (1 + \frac{\tU}{\pi v_{F}})$ and
$v_{\sigma}/K_{\sigma} = v_{F} (1 - \frac{\tU}{\pi v_{F}})$.  
Among the various cosine potentials, 
we have the usual spin-flip backscattering $\cos(\sqrt{8}\phi_{\sigma})$ and 
Umklapp $\cos(\sqrt{8}\phi_{\rho})$ terms as well as the triplet 
superconducting $\cos(\sqrt{2}\theta_{\rho})\cos(\sqrt{2}\theta_{\sigma})$ 
term. The chemical potential term can be absorbed 
by performing the shift $\phi_{\rho}\rightarrow \phi_{\rho} 
+ \frac{\sqrt{2}K_{\rho}\mu}{v_{\rho}} x$. 
The cosine potentials with couplings $V_{1}$ and $V_{2}$ 
are generated under RG by the $\tp$ and $\tV$ terms. 
Using the operator product expansion~\cite{giamarchi}, 
we find the RG equations for the various couplings to second-order as 
\bea
\frac{d U_{\rho}}{dl} &=& (2 - 2K_{\rho}) U_{\rho}\nonum\\
\frac{d U_{\sigma}}{dl} &=& (2 - 2K_{\sigma}) U_{\sigma} 
- (\frac{1}{K_{\sigma}} - K_{\sigma})\tp^{2}\nonum\\
\frac{d \tV}{dl} &=& (2 - \frac{1}{2}(\frac{1}{K_{\sigma}} 
+ \frac{1}{K_{\rho}})) \tV - K_{\sigma}\tp V_{2} \nonum\\
\frac{d\tp}{dl} &=&\hspace*{-0.2cm} (2 - \frac{1}{2}(K_{\sigma} 
+ \frac{1}{K_{\sigma}})) \tp \hspace*{-0.1cm}-\hspace*{-0.1cm}
\frac{\tV V_{2}}{K_{\rho}} \hspace*{-0.1cm}-\hspace*{-0.1cm}
(K_{\sigma} U_{\sigma}+\frac{V_{1}}{K_{\sigma}})2\tp\nonum\\
\frac{d V_{1}}{dl} &=& (2 - \frac{2}{K_{\sigma}}) V_{1} 
+ (\frac{1}{K_{\sigma}} - K_{\sigma})\tp^{2}\nonum\\
\frac{d V_{2}}{dl} &=& (2 - \frac{1}{2}(K_{\sigma} 
+ \frac{1}{K_{\rho}})) V_{2} - \frac{\tp\tV}{K_{\sigma}}~,
\label{rgeq1}
\eea
where all couplings have been normalised with respect to the 
quantity $2\pi v_{F}$, and we have set $2\pi v_{F} = 1$ for notational 
simplicity.
The RG equations for the two interaction parameters $(K_{\rho}, K_{\sigma})$,
the two velocities $(v_{\rho}, v_{\sigma})$ 
as well as the parameter $\delta=K_{\rho}\mu/v_{\rho}$ are found to be 
\bea
\frac{d K_{\sigma}}{dl} &=& 
-K_{\sigma}^{2} (U_{\sigma}^{2} + V_{2}^{2} + \tp^{2}) 
+ V_{1}^{2} + \tp^{2} + \tV^{2}\nonum\\
\frac{d K_{\rho}}{dl} &=& -K_{\rho}^{2} U_{\rho}^{2}
J_{0}(\delta (l)\alpha) + V_{2}^{2} + \tV^{2}\nonum\\
\frac{d v_{\sigma}}{dl} &=&
-v_{\sigma} K_{\sigma} (U_{\sigma}^{2} + V_{2}^{2} + \tp^{2}) 
+ \frac{v_{\sigma}}{K_{\sigma}} (V_{1}^{2} + \tp^{2} + \tV^{2})\nonum\\
\frac{d v_{\rho}}{dl} &=&
-v_{\rho} K_{\rho} U_{\rho}^{2}J_{2} (\delta (l)\alpha) 
+ \frac{v_{\rho}}{K_{\rho}} (V_{2}^{2} + \tV^{2})\nonum\\
\frac{d\delta}{dl}&=& \delta (l) - U_{\rho}^{2}J_{1} (\delta (l)\alpha)~,
\label{rgeq2}
\eea
where $\delta (l) = \delta e^{l}$, 
$\alpha$ is a short-distance cut-off like the lattice spacing and 
$J_{0}(x), J_{1}(x)$ are Bessel functions~\cite{giamarchi}. The various 
second order correction terms arise from $(\partial_{x}\phi_{\rho/\sigma})^{2}$ 
and $(\partial_{x}\theta_{\rho/\sigma})^{2}$ that are generated under the 
RG transformations~\cite{giamarchi,hocazgia,bencasgia}. As discussed in 
Ref.(\cite{bencasgia}), the competing influences of the various couplings 
on the renormalisations of the interaction parameters $(K_{\rho},K_{\sigma})$ 
can cause their values to either grow or decrease. In turn, this affects 
drastically the scaling dimensions of the various couplings and can lead to 
the system undergoing a transition from one type of ordered phase (in which 
a particular coupling grows the fastest to strong coupling) to another, 
with the passage being through a gapless (critical) phase. 
The existence of such a critical (gapless) region in coupling space, lying 
in between two ordered (massive) phases, is revealed in a subsequent analysis. 
Further, such a deconfined phase can be thought of as the quasi-1D analog of 
that analysed in detail later via a RPA treatment when dealing with many coupled 
chains.

\subsection{Phase Diagram for repulsive inter-TFIM coupling}
For repulsive interactions ($U_{\perp}>0$) between the bond-fermions, 
the $\sigma$ sector is massless and $K_{\sigma}$ flows 
under RG to the fixed point value $K_{\sigma}^{*} \gtrsim 1$ 
and $1/2\leq K_{\rho}\leq 1$. At 1/2-filling (for the bond-fermions), 
the couplings $U_{\rho}$, 
$\tV$, $\tp$, $V_{1}$ and $V_{2}$ are all relevant while $U_{\sigma}$ 
is irrelevant. The competition to reach strong-coupling first is, 
however, mainly between $U_{\rho}$, $\tp$ and $\tV$.
We show below the phase diagram as derived from this analysis.
\begin{figure}[htb]
\begin{center}
\scalebox{1}{
\psfrag{1}[bl][bl][0.9][0]{$K_{\rho}>(1/K_{\sigma},1/4(K_{\sigma}+1/K_{\sigma}))$}
\psfrag{2}[bl][bl][0.9][0]{$1/K_{\sigma}>K_{\rho}>1/4(1/K_{\sigma} + /K_{\rho})$}
\psfrag{3}[bl][bl][0.9][0]{$K_{\sigma}<(1/4(1/K_{\rho}+1/K_{\sigma}),1/4(K_{\sigma} + 1/K_{\sigma}))$}
\includegraphics{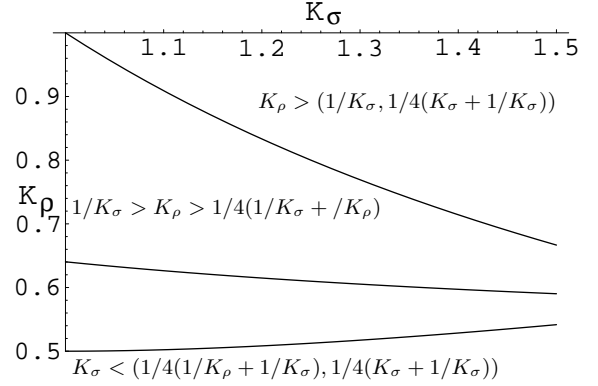}
}
\end{center}
\caption{The RG phase diagram in the $(K_{\sigma}, K_{\rho})$ plane
for repulsive interchain interactions ($U_{\perp}<0$). 
The three regions $K_{\rho}<(1/4(1/K_{\sigma} + 1/K_{\rho}), 
1/4(K_{\sigma} + 1/K_{\sigma}))$, 
$1/K_{\sigma}>K_{\rho}>1/4(1/K_{\sigma} + 1/K_{\rho})$ and 
$K_{\rho}>(1/K_{\sigma}, 1/4(K_{\sigma} + 1/K_{\sigma}))$ give the 
values of $(K_{\sigma},K_{\rho})$ for which the 
couplings $U_{\rho}$, $\tp$ and $\tV$ respectively are the fastest 
to grow under RG.}  
\label{phased1}
\end{figure}
\par
In the phase diagram in Fig.(\ref{phased1}), the three lines with intercepts 
at $(K_{\sigma}=1, K_{\rho}=1)$, $(K_{\sigma}=1, K_{\rho}=0.64)$ and 
$(K_{\sigma}=1, K_{\rho}=1/2)$ are the relations $K_{\sigma}=1/K_{\rho}$, 
$K_{\rho}=1/4(1/K_{\sigma} + 1/K_{\rho})$ and 
$K_{\rho}=1/4(K_{\sigma} + 1/K_{\sigma})$ respectively. The regions 
$K_{\rho}<(1/4(1/K_{\sigma} + 1/K_{\rho}), 1/4(K_{\sigma} + 1/K_{\sigma}))$, 
$1/K_{\sigma}>K_{\rho}>1/4(1/K_{\sigma} + 1/K_{\rho})$ and 
$K_{\rho}>(1/K_{\sigma}, 1/4(K_{\sigma} + 1/K_{\sigma}))$ signify 
the values of $K_{\rho}$ and $K_{\sigma}$ for which $U_{\rho}$ (in-chain 
Wigner charge-ordered Mott insulator), $\tp$ and $\tV$ (in-chain 
Peierls charge-ordered Mott insulator of preformed bond-fermion pairs) 
respectively are the fastest to reach strong-coupling. The RG equations 
for the coupling $U_{\rho}$, the interaction parameter $K_{\rho}$ and 
the incommensuration parameter $\delta$ are familiar from the literature 
on commensurate-incommensurate transitions~\cite{japaridze}. For 
temperatures $T>>K_{\rho}\mu$, the finite chemical potential (arising 
from the non-zero transverse-field strength $t$) is unable 
to quench the Umklapp scattering processes, allowing for the growth of 
$U_{\rho}$ to strong-coupling. For $T<<K_{\rho}\mu$, the finite chemical 
potential cuts off the RG flow of $U_{\rho}$, freezing the Umklapp 
scattering processes. 
\par
For the case of $\tp$ being the most relevant coupling, 
we find our RG equations to be a non-trivial 
generalisation of those derived in~\cite{kusmartsev} for the model of 
two coupled spinless fermion chains without the intrachain $\tV$ 
pairing term. The resulting picture then 
describes strong interchain two-particle correlations between bond-fermions 
sharing a rung. Following the analysis 
outlined in~\cite{kusmartsev,tsvelik}, we conclude that this phase is 
a novel insulating phase characterised by a mass gap and delocalised 
bond-fermions on rungs, resembling the orbital antiferromagnetic ground 
state found in the spinless two-chain problem. This matches our finding of 
an orbital antiferromagnetic ground state in the strongly-coupled ladder 
with dominant antiferromagnetic rung-couplings in an earlier 
work~\cite{first}.  
Away from 1/2-filling (for the bond-fermions), the competition is mainly 
between $\tp$ and $\tV$. For $\tV$ reaching strong-coupling ahead 
of $\tp$, the system is in a channel triplet-spin singlet superconducting 
phase with mobile intra-chain hole pairs; for $\tp$ reaching 
strong-coupling ahead of $\tV$, we are currently unable to describe in 
more detail the dominant instability away from the 
orbital antiferromagnetism like insulating phase away from quarter-filling.  
\subsection{Gapless phase driven by inter-TFIM hopping}
\begin{figure}[htb]
\begin{center}
\scalebox{0.4}{
\psfrag{1}[bl][bl][3][0]{$V_{1}$}
\psfrag{2}[bl][bl][3][0]{$\tp$}
\psfrag{3}[bl][bl][3][0]{I}
\psfrag{4}[bl][bl][3][0]{II}
\psfrag{5}[bl][bl][3][0]{$(0,0)$}
\includegraphics{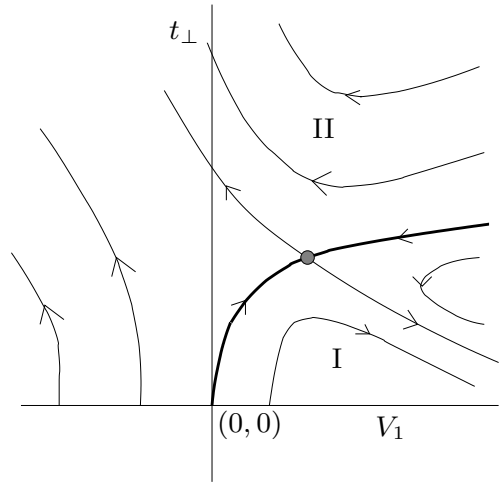}
}
\end{center}
\caption{The RG phase diagram in the $(V_{1}, \tp)$ plane. The thick line 
characterises the set of points which flow to the intermediate fixed point 
$(V_{1}^{*}, \tp^{*})$ shown by the filled circle. The thin lines show 
all RG flows which flow towards strong-coupling in the two phases {\bf I} 
and {\bf II}, characterised by the relevant couplings $\tV$ and $\tp$ 
respectively.} 
\label{intfix}
\end{figure}

Interestingly, while studies of ladder models 
have shown a plethora of charge and spin-ordered gapped phases
~\cite{giamarchi,donohue,orignac}, there remains 
the intriguing possibility that some of these gapped phases may 
themselves be separated from one another by non-trivial gapless phases 
of finite width~\cite{tsvelik}. In what follows, we provide an 
explicit realisation of this scenario. 
The RG equations (\ref{rgeq1})-(\ref{rgeq2}) reveal the existence of 
a non-trivial fixed point (FP) for any value of $(K_{\rho},K_{\sigma})$ 
lying the in ranges $1/2<K_{\rho}<1$, $1<K_{\sigma}<2+\sqrt{3}$ and which 
is perturbatively accessible from the trivial weak-coupling FP. 
While perturbative RG is known to have its limitations in such cases, the 
finding of a non-trivial FP is nevertheless reliable as long as the values 
of the various couplings at the FP are small (as indicated 
above). We note 
that a similar non-trivial fixed point was found in a study of the 
anisotropic Heisenberg spin-1/2 chain in a magnetic field~\cite{duttasen}, 
where the authors derived a set of RG equations which were very similar 
to those found in~\cite{kusmartsev}. Here, the non-trivial FP is given by
\bea
\tp^{*} &=& \sqrt{ab}~,~V_{1}^{*} = \frac{K_{\sigma}^{*}+1}{2}{\tp^{*}}^{2}~,~
U_{\sigma}^{*} = \frac{V_{1}^{*}}{2K_{\sigma}^{*}}\nonum\\
\tV^{*} &=& \sqrt{aK_{\rho}^{*}(cK_{\sigma}^{*} - (K_{\sigma}^{*}+1)^{2}ab)} 
~,~ V_{2}^{*} = \sqrt{\frac{b}{a}}\frac{\tV^{*}}{K_{\sigma}^{*}}
\eea 
where
$a = 2-(K_{\sigma}^{*} + 1/K_{\rho}^{*})/2$~,~
$b = 2-(1/K_{\sigma}^{*} + 1/K_{\rho}^{*})/2$ and
$c = 2-(K_{\sigma}^{*} + 1/K_{\sigma}^{*})/2$~.
Further, we can safely make the approximation of the renormalisations of 
$K_{\rho}$ and $K_{\sigma}$ being small at this non-trivial FP 
~\cite{duttasen}. The system is 
gapless at this non-trivial FP as well as at points which flow to it. 
The trivial FP has 6 unstable directions ($U_{\rho},\tV,\tp,V_{1},V_{2}$ 
and $\delta$), 1 stable directions ($U_{\sigma}$) 
and 2 marginal 
directions ($K_{\rho}$ and $K_{\sigma}$). The non-trivial FP has 5 unstable 
directions, 2 stable directions and two marginal directions. 
The presence of the two
stable directions at the non-trivial FP indicates the existence of a 
two-dimensional surface of gapless theories in the five-dimensional 
$(U_{\sigma},\tV,\tp,V_{1},V_{2})$ coupling space. This gapless phase is 
the analog of the ``Floating Phase" found in the phase diagram of the 
1D axial nnn Ising model~\cite{duttasen}. We 
present in Fig.(\ref{intfix}) below a RG flow phase diagram which 
is projected onto the 
$(V_{1}, \tp)$ plane (a similar RG flow diagram is found for the case 
of the anisotropic Heisenberg model in a magnetic field $h$~\cite{duttasen} 
in the $(a,h)$ plane, where $a$ is the anisotropy parameter). 
\par
The regions {\bf I} and {\bf II} characterise all RG flows 
which do not flow to the intermediate fixed point at 
$(V_{1}^{*}, \tp^{*})$. The transitions from the non-trivial FP towards 
either of phases {\bf I} and {\bf II} fall under the universality 
class of the Kosterlitz-Thouless type~\cite{duttasen}. 
In region {\bf I}, $V_{1}$ flows to 
strong-coupling while $\tp$ decays; for $1/2<K_{\rho}<1$ and 
$K_{\sigma}>1$, we know from the above discussion that in this region, 
the coupling $\tV$ will reach strong-coupling first. In region {\bf II}, 
both $\tp$ and $V_{1}$ grow under RG, with the coupling $\tp$ being the 
first to reach strong-coupling. Thus, the RG trajectory leading to the 
intermediate fixed point represents a gapless phase separating the 
two gapped, charge-ordered phases {\bf I} and {\bf II} characterised 
by the relevant couplings $\tV$ and $\tp$ respectively. Nonperturbative 
insight on such critical phases is also gained in section IV, when 
we treat the case of many such TFIM systems coupled to one another using
the RPA method. 
\subsection{Phase diagram for attractive inter-TFIM coupling}
For attractive interactions ($U_{\perp}<0$) between the 
bond-fermions, we can carry out a similar analysis. In this case, 
we can see that $K_{\rho}>1$ while $K_{\sigma}<1$. 
Then, from the RG equations given above, we can see that the Umklapp coupling 
$U_{\rho}$ and $V_{1}$ are irrelevant while the 
couplings $\tp$, $U_{\sigma}$, $\tV$ and $V_{2}$ 
are relevant. The competition to reach strong-coupling first is, however, 
mainly between $U_{\sigma}$, $\tp$ and $\tV$. 
We show below the phase diagram at 1/2-filling for the bond-fermions 
as derived from this analysis.   
\begin{figure}[htb]
\begin{center}
\scalebox{1}{
\psfrag{1}[bl][bl][1][0]{$K_{\sigma}>(1/K_{\rho},1/4(1/K_{\rho}+1/K_{\sigma}))$}
\psfrag{2}[bl][bl][1][0]{$1/K_{\rho}>K_{\sigma}>1/\sqrt{3}$}
\psfrag{3}[bl][bl][1][0]{$K_{\sigma}<(1/4(1/K_{\rho}+1/K_{\sigma}),1/\sqrt{3})$}
\includegraphics{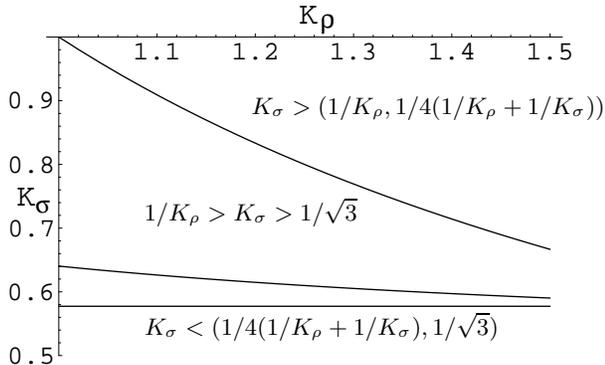}
}
\end{center}
\caption{The RG phase diagram in the $(K_{\rho}, K_{\sigma})$ plane 
for attractive interchain interactions ($U_{\perp}>0$). The three regions 
$K_{\sigma}<(1/4(1/K_{\sigma} + 1/K_{\rho}), 
1/\sqrt{3})$, $1/K_{\rho}>K_{\sigma}>1/\sqrt{3}$ and 
$K_{\sigma}>(1/K_{\rho}, 1/4(1/K_{\rho} + 1/K_{\sigma}))$
give the values of $(K_{\sigma},K_{\rho})$ for which the 
couplings $U_{\sigma}$, $\tp$ and $\tV$
respectively are the fastest to grow under RG.
} 
\label{phased2}
\end{figure}
\par
In the phase diagram in Fig.(\ref{phased2}), the three lines with intercepts 
at $(K_{\rho}=1, K_{\sigma}=1)$, $(K_{\rho}=1, K_{\sigma}=0.64)$ and 
$(K_{\rho}=1, K_{\sigma}=1/\sqrt{3})$ are the relations 
$K_{\sigma}=1/K_{\rho}$, 
$K_{\sigma}=1/4(1/K_{\sigma} + 1/K_{\rho})$ and $K_{\sigma}=1/\sqrt{3}$ 
respectively. The regions $K_{\sigma}<(1/4(1/K_{\sigma} + 1/K_{\rho}), 
1/\sqrt{3})$, $1/K_{\rho}>K_{\sigma}>1/\sqrt{3}$ and 
$K_{\sigma}>(1/K_{\rho}, 1/4(1/K_{\rho} + 1/K_{\sigma}))$ signify 
the values of $K_{\rho}$ and $K_{\sigma}$ for which $U_{\sigma}$ (rung-dimer 
insulator with in-chain Wigner charge-ordering), $\tp$ 
and $\tV$ (insulator with in-chain dimers and Peierls charge-ordering) 
respectively are the fastest to reach strong-coupling.
This matches our finding of a ground state with in-chain Wigner charge 
order and rung-dimers in the strongly-coupled ladder 
with large ferromagnetic rung-couplings in an earlier 
work~\cite{first}. 
Away from 1/2-filling (for the bond-fermions), depending on which of 
the three couplings $\tp$, $\tV$ and $U_{\sigma}$ is the first to reach 
strong-coupling, the system exists either as a superconductor 
with intra-chain hole pairs ($\tV$) or a superconductor with 
rung-singlet hole pairs ($U_{\sigma}$) or a phase reached by following 
the dominant instability away from the orbital antiferromagnetism like 
insulating phase ($\tp$) but which we are currently unable to describe 
in greater detail.
\section{Coupled TFIM Systems: quantum criticality, dimensional crossover 
and deconfinement}
Having studied the rich phase diagram of a 2-leg coupled TFIM system in the 
preceeding section in considerable detail, we now proceed to investigate the 
case when many such TFIM systems are coupled to one another. This is done 
with an effort towards gaining an understanding of how such coupled systems 
undergo a dimensional crossover from nearly isolated quasi-one dimensional 
TIFM systems to a anisotropic strongly coupled system in higher dimensions. 
Dimensional crossover in coupled spin systems has been studied using 
renormalisation group (RG) arguments~\cite{affleck,giamarchi} and the 
random phase approximation (RPA) for Ising chains~\cite{scalapino,carr2}, 
Heisenberg chains~\cite{schulz,bocquet,giamarchi} and ladders in a magnetic
field~\cite{giatsvelik} and ladders with frustration~\cite{starykh}. 
Having used perturbative RG arguments in exploring the 
phase diagram of the 2-leg system earlier, we now employ the RPA method to 
study the passage to higher dimensionality. This is in keeping with the fact 
that dimensional crossover is essentially a nonperturbative phenomenon
~\cite{giamarchi}. At the same time, while the mean-field-like approach 
of RPA is exact only in infinite dimensions (i.e., infinite coordination 
number), its application to the physics of coupled quasi-1D spin systems 
for small coordination numbers (i.e., lower dimensions) has met with success
~\cite{irkhin,bocquet2}. It is also worth noting that while a naive mean-field 
treatment of single particle-hopping between fermionic chains is not possible 
as a single fermion operator has no well-defined classical limit
~\cite{giamarchi}, we are able to treat two-particle hopping processes between 
our underlying chain systems~\cite{first} (or, ladder 
systems~\cite{mostovoy,fulde,horsch}) systems via RPA by working with an 
effective theory in terms of pseudospins (as evidenced by the inter chain bond 
fermion hopping processes studied earlier for the case of the 2-leg ladder).
This is justified because the presence of the large on-site Hubbard coupling 
in our underlying model makes the single particle hopping irrelevant (in a 
RG sense) while two-particle processes (including hopping terms) can be 
crucial in determining the phase diagram~\cite{giamarchi}. A full treatment 
including single-particle hopping will require a treatment using 
chain-dynamical mean field theory (c-DMFT)~\cite{arrigoni} and will be the 
focus of a future work. In what follows, we will follow instead the RPA 
method outlined in Ref.\cite{carr2}. 
\par
We treat the dynamics of the coupled spin system for the two inter-TFIM 
couplings, $J_{\perp}$ and $t_{\perp}$, in eq.(\ref{twochain}) given above 
using the RPA method in turn. Beginning with the coupling $t_{\perp}$, 
this method involves computing the dynamical spin susceptibility $\chi$ 
of the coupled system in the disordered phase as 
\beq
\chi (\omega, k, \vec{k}_{\perp}) = 
[\chi_{1D}^{-1}(\omega,k)-t_{\perp}(\vec{k}_{\perp})]^{-1} 
\label{rpamain}
\eeq
in terms of the frequency $\omega$, the longitudinal and transverse wave-vectors 
$k$ and $\vec{k}_{\perp}$ respectively. $\chi_{1D}$ is the dynamical spin 
susceptibility of a single TFIM system, to be calculated assuming incipient 
order along the $\tau^{x}$ direction in pseudospin space 
\beq
\chi_{1D}(\omega,k) = -i\sum_{n}\int_{0}^{\infty}dt e^{i(\omega t - kn)}
\langle\lbrack\tau^{x}(t,n),\tau^{x}(0,0)\rbrack\rangle
\label{1dchi}
\eeq
and the transverse coupling $t_{\perp}(\vec{k}_{\perp})
\sim z_{\perp}t_{\perp}(\vec{k}_{\perp}=0)$, for each TFIM system having 
a coordination number of $z_{\perp}$. Then, a divergence in the 
dynamical pseudospin correlation function $\chi (\omega,k,\vec{k_{\perp}})$ 
signifies an instability towards the formation of an ordered state. However, 
before setting out with the calculations, it is worth pausing to consider 
first the likely effects of a transverse coupling like $t_{\perp}$. As 
discussed earlier, the phase diagram at $T=0$ and $t_{\perp}=0$ is simple, 
with an ordered phase for $2t<V$ ($g=(2t-V)/V<0$), a quantum disordered phase 
for $2t>V$ ($g>0$) and a quantum critical point at $2t=V$ ($g=0$). A 
finite $t_{\perp}$ will cause 
the ordered phase to be extended to finite temperatures (with a critical 
temperature $T_{c}$ for the case of $2t=V$), with a first order 
phase boundary ending at a new quantum critical point (QCP) $g_{c}=\Delta_{c}/V$ at 
$T=0$~\cite{carr2}. As for the simple TFIM~\cite{sachdev}, there exists a 
``quantum critical" region just above the QCP and to the right of the ordered 
phase, with a crossover to the disordered phase at finite $T$. This is shown 
schematically in the $T-g$ phase diagram given below in Fig.(\ref{qcp}). The 
transition belongs to the 3D Ising universality class while the QCP to the 
4D Ising universality class~\cite{itzdrouffe}.
\begin{figure}[htb]
\begin{center}
\scalebox{0.35}{
\psfrag{1}[bl][bl][3.5][0]{$T_{c}$}
\psfrag{2}[bl][bl][4]{$T$}
\psfrag{3}[bl][bl][3.5][0]{$0$}
\psfrag{4}[bl][bl][3.5][0]{$g_{c}$~({\bf QCP})}
\psfrag{5}[bl][bl][4][0]{$g$}
\psfrag{6}[bl][bl][3][0]{Ordered}
\psfrag{7}[bl][bl][3][0]{Gapless}
\psfrag{8}[bl][bl][3][0]{Disordered}
\psfrag{9}[bl][bl][3][0]{Phase}
\psfrag{10}[bl][bl][3][0]{Phase}
\psfrag{11}[bl][bl][3][0]{Critical}
\psfrag{12}[bl][bl][3][0]{Region}
\psfrag{13}[bl][bl][3][0]{Quantum}
\includegraphics{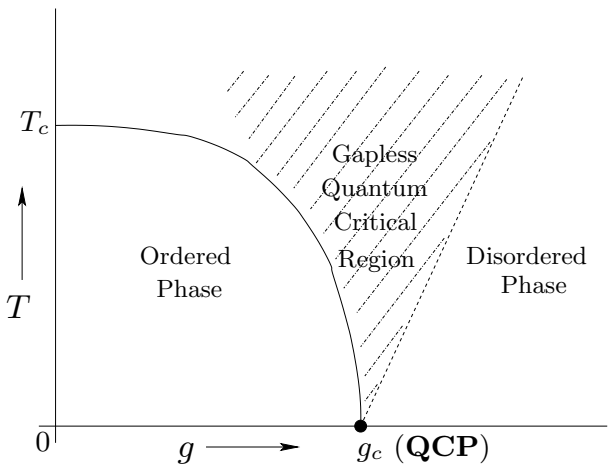}
}
\end{center}
\caption{The $T-g$ phase diagram for the case of many TFIM systems coupled 
by a transverse coupling $t_{\perp}$ (or $J_{\perp}$). The $T=0$ Ordered 
Phase of the uncoupled TFIM is now extended, with a phase boundary which 
has a value $T_{c}$ for the model at $g=0$ and a $T=0$ quantum critical 
point (QCP) at $g\equiv g_{c}$. The hashed region immediately to the right 
of the ordered phase and just above the QCP is the gapless quantum critical 
region (described in the text). The dashed line represents a finite $T$ 
crossover from the quantum critical region to a Disordered phase.} 
\label{qcp}
\end{figure}
\par
In the following, we compute, via the RPA method,  
the quantities $g_{c}$ at $T=0$, $T_{c}$ for the case of $g\equiv\Delta/V=0$, 
the shape of the phase boundary near the QCP, the dynamical spin susceptibility 
$\chi(\omega,k,\vec{k}_{\perp})$ at the QCP as well as the dispersion in the 
transverse directions for small $t_{\perp}$ and close to the QCP~\cite{carr2}. 
We focus our attention mainly on, and in the neighbourhood of, the QCP in order to 
stress the role played by the critical quantum fluctuations in determining 
the physics of deconfinement and dimensional crossover in our system. In 
this, we are aided by the integrability and conformal invariance of the 
TFIM model for $2t=V$,~$t_{\perp}=0$~\cite{lieb,niemeijer,sachdev}; this allows 
us to exploit the nonperturbative results for a single TFIM system 
(as long as $\Delta=|2t-V|<<1$), while dealing with the physics of the 
transverse couplings at a mean-field level. At the same time, the spectrum 
and dispersion deep inside the ordered phase also 
proves to be fascinating~\cite{carr2}. Specifically, it has been 
demonstrated that far from the transition line, dispersion in the 
transverse directions is very weak (i.e., the spectrum nearly 
one-dimensional), and a hidden $E_{8}$ symmetry of the underlying exactly 
solvable model~\cite{zamol} gives rise to a spectrum of eight massive 
particles (three of which should be experimentally observable). 
\par
We begin by computing the critical value of the transverse coupling 
$g_{c}$ at $T=0$. For this, we can use the expression for slightly 
off-critical $\chi_{1D}$ for the case when the mass of the 
single TFIM system is very small ($m=\Delta_{c} << 1$). This, for small 
$\omega$, is given by   
\beq
\chi_{1D}(\omega,k) \simeq \frac{Z_{0} V (\Delta_{c}/V)^{1/4}}
{\omega^{2}-(v k)^{2} - \Delta_{c}^{2}}
\label{offcrit1dchi}
\eeq
where the velocity $v=V\alpha$ (where $\alpha$ is the lattice spacing) and 
$Z_{0}=1.8437$. Then, from the divergence of the susceptibility of the coupled 
system, $\chi(\omega,k,\vec{k}_{\perp})$ 
\beq
\chi_{1D}^{-1}(\omega,k) = z_{\perp}t_{\perp}(\vec{k}_{\perp}=0),
\label{diverge}
\eeq 
we get, for the case of $\omega=0=k$,
\beq
\frac{V}{Z_{0}} g_{c}^{7/4}\approx z_{\perp}t_{\perp}
\eeq
where we've dropped the argument of $\vec{k}_{\perp}=0$ in $t_{\perp}$ 
for the sake of convenience. Thus, we get
\beq
g_{c}\approx c_{1}(\frac{z_{\perp}t_{\perp}}{V})^{4/7}
\label{critg}
\eeq
where the constant $c_{1}=Z_{0}^{4/7}=1.42$. Precisely the same expression for the 
mass in the ordered phase and very close to the QCP, $\Delta=g_{c}V$, is also obtained 
by carrying out a self-consistent treatment of the effective magnetic field, 
$h=z_{\perp}t_{\perp}\langle\tau^{x}\rangle$ in the TFIM for a single chain.
For this, one uses the slightly off-critical susceptibility 
for the 1D TFIM given earlier (eqn.(\ref{offcrit1dchi})) with the mass 
$\Delta$ replaced by $\Delta(1+(h/V)^{2})$~~\cite{delfino}. From this result,
the authors of Ref.(\cite{carr2}) concluded that  the dispersion in the transverse 
directions in the ordered phase and close to the QCP is much stronger than that 
deep in the ordered phase.   
\par
To calculate the $T_{c}$ for 
the case of $g=0$, we use the $\chi_{1D}(\omega=0,k=0)$ at finite $T$
\beq
\chi_{1D}(\omega=0,k=0) = \frac{c_{2}}{V}(\frac{2\pi T}{V})^{-7/4}
\label{omega0k01dchi}
\eeq 
where the constant $c_{2}=\sin(\pi/8){\rm B}^{2}(1/16,7/8)$ and ${\rm B}(x,y)$ 
is the Euler Beta function. Thus, we find, by using eqn. (\ref{diverge})
\beq
\frac{T_{c}}{V} = c_{3}(\frac{z_{\perp}t_{\perp}}{V})^{4/7}
\label{critT}
\eeq
where the constant $c_{3}= c_{2}/(2\pi) = 2.12$. Further, we can also 
compute the susceptibility $\chi$ for the coupled system at the QCP for small 
$\vec{k}_{\perp}$ by using the relation for the slightly off-critical 
$\chi_{1D}$ given earlier, eqn.(\ref{offcrit1dchi}), together with the relation 
$\Delta_{c}^{2}=g_{c}^{1/4}Vt_{\perp}$ in eqn.(\ref{rpamain}), giving 
\beq
\chi (\omega,k,\vec{k}_{\perp})\sim\frac{Z_{O} V g_{c}^{1/4}}
{\omega^{2} - (v k)^{2} - (\vec{v}_{\perp}\cdot\vec{k}_{\perp})^{2}},
\label{critchi}
\eeq
where $|\vec{v}_{\perp}|^{2}= (Z_{0} V g_{c}^{1/4}/2)
d^{2}t_{\perp}(\vec{k}_{\perp}=0)/d\vec{k}_{\perp}^{2}$ is gained 
by a Taylor expansion to second order~\cite{carr2}. Further, the shape 
of the phase boundary at low $T$ can be determined by using the 
$\chi_{1D}$ of the TFIM at low $T$~~\cite{sachdev}
\beq
\chi_{1D} (\omega, k) = \frac{Z_{0}(\alpha\Delta/v)^{1/4}}
{(\omega + i/\tau_{\psi})^{2} - (v k)^{2} - \Delta^{2}}, 
\label{offcritfiniteTchi}
\eeq
where $\tau_{\psi}=\frac{\pi}{2T}e^{\Delta/T}$ is the dephasing time 
due to quantum fluctuations. Then, for $\omega=0=k$ in $\chi_{1D}$, 
the eqn.(\ref{diverge}) gives
\beq
\ln T - \frac{\Delta}{T} = \ln m + \ln \Lambda,
\label{transcen}
\eeq
where $\Lambda=\frac{\pi}{2}(\frac{Z_{0}t_{\perp}}{g^{7/4}V} - 1)^{1/2}$.
The expression (\ref{transcen}) given above has an approximate solution~\cite{carr2}
\beq
T_{c} = \frac{\Delta}{\ln(1/\Lambda) - \ln\ln(1/\Lambda)}~.
\label{phaseboundary}
\eeq
This relation gives us the shape of the phase boundary for low $T$ and close to 
the QCP. In this way, we have derived the key features of the $T-g$ phase diagram 
for the case of the $t_{\perp}$ transverse coupling (Fig.(\ref{qcp})) given above. 
We can also carry out an identical RPA calculation for the other transverse coupling, 
$J_{\perp}$, for the case when the single chain system is critical, 
$2t=V, t_{\perp}=0$ in equn.(\ref{twochain})~\cite{carr2}. This is because this
theory is again known to be integrable~\cite{zamol} and falls in the same universality 
class as that of the TFIM. We can thus determine an identical set of relations for 
(i) the critical 
coupling $g_{c}$ at $T=0$, (ii) the critical temperature $T_{c}$ for the case of $g=0$, 
(iii) the susceptibility $\chi$, (iv) dispersion in the transverse directions as 
well as (v) the shape of the phase boundary close to the QCP to those obtained 
earlier, but with $t_{\perp}$ replaced by $J_{\perp}$ everywhere. In this way, 
we obtain essentially the 
same $T-g$ phase diagram for the case of the $J_{\perp}$ transverse coupling, with 
the only difference being the fact that the spectrum of the ordered phase is now 
obtained from the solution of the critical 1D TFIM in a longitudinal 
field~\cite{zamol}.   
\par
We can see from the results given above that, as we move along the interior of the 
ordered phase towards the phase boundary, the excitation gaps decrease together with 
a gradual growth of the dispersion in the transverse directions. At the QCP, the 
spectrum is gapless and this is reflected in the fact that the dynamical 
susceptibility $\chi$ at the QCP can have at most logarithmic corrections. At 
low $T$, directly in the region above the QCP, the spectrum and dynamics are 
mainly governed by the quantum critical point while the thermal excitations 
are described by the associated continuum quantum field theory. By this, we mean 
that for energy scales given by $\omega >> k_{B}T$, the system still resembles 
a quantum critical one in this region of the phase diagram while for 
$\omega << k_{B}T$, a relaxation of the dynamics (given by $\tau_{\psi}^{-1}$) 
caused by quantum fluctuations sets in. The rapid growth of the dispersion 
in the transverse directions close to the QCP is the physics of the {\it 
dimensional crossover} in this system, while the vanishing of all mass  
gaps at the QCP is the physics of {\it deconfinement}. These 
results echo our earlier finding of the critical phase in the phase diagram of the 
2-leg TFIM system by starting from the $\tilde{V}$ ordered phase and increasing 
$t_{\perp}$. While the integrability of the TFIM allows us to make considerable 
progress in computing various quantities, the role of the QCP in the mechanism 
responsible for the dimensional crossover and deconfinement appear to be robust. 
These results lead us, therefore, to the conclusion that critical 
quantum fluctuations associated with a QCP can quite generically cause such critical 
(gapless) phases to emerge in higher dimensions from the ordered (gapped) phases 
of lower dimensional systems when coupled to one another.   

\section{Comparison with recent numerical works}
We present here a brief discussion of the relevance of our work to some numerical 
investigations that have been carried out on 2-leg ladder systems as well as 
coupled TFIM systems. Vojta et al.~\cite{vojta} studied the problem of a 
strongly correlated electronic problem at $1/4$-filling and with extended 
Hubbard interactions (keeping only a nn repulsion) using the 
DMRG method. The phase diagram they obtained contains several phases with 
charge and/or spin excitation gaps. While a comparison of our work with this 
study is hindered by the fact that the DMRG analysis does not have the crucial 
element of the nnn repulsion ($V_{2}$ in our work), Fig.2 of 
that work reveals that for the case of $U>>V_{1}>t$, the authors indeed find a 
charge-ordered CDW state (i.e, the Wigner charge ordered state of 
Ref.(\cite{first})) with an excitation gap in the spin sector as well. This is 
in conformity with our finding of a Wigner charge-ordered state with a spin 
gap for the case of the $U_{\rho}$ coupling being the most relevant under RG. 
Further, the $t_{\perp}$ of that study corresponds to the single-particle 
hopping between the legs while our work has focused on the effects of 
two-particle hopping. Finally, with the on-site Hubbard coupling, $U$, being 
the largest in the problem, we are unable to see any phase-separated state 
in our phase diagram (as observed by Vojta et al.). 
\par 
In a slightly earlier work, Riera et al.~\cite{riera} studied the case of 
$1/4$-filled chain/ladder Hubbard and $t-J$ systems, but which also include Holstein 
and/or Peierls-type couplings to the underlying lattice. Their findings reveal 
coexisting charge and spin orders in both chain and ladder systems. Specifically, 
for the case of their chain system, by keeping only on-site and nn 
repulsion together with an on-site Holstein-type coupling of the electronic 
density to a phonon field, their phase diagram (Figs.4) reveal separate phases 
with Wigner as well Peierls-type charge order. This is in keeping with our 
findings, but the origin of the Peierls order in the two cases are different: 
in our work, it originates from the competition of the nnn coupling $V_{2}$ with 
the nn coupling $V_{1}$, while in their work, it needs the Holstein coupling to 
the lattice. Riera et al. find a similar phase diagram (Figs.5) for the case 
of an extended $t-J$ model (i.e., including nnn t and J couplings) with a  
Holstein coupling. The addition of a Peierls-type coupling leads to a 
spin-Peierls instability, i.e., the formation of a dimerised spin order, 
which coexists with the Peierls-type charge order. Again, while this matches our 
findings, the origins are different. Qualitatively similar conclusions are also 
reached by the authors in their study of an anisotropic 2-leg $t-J$ ladder with 
an extended on-chain nn coupling and Holstein/Peierls-type lattice couplings 
(Figs.2 and 3).
\par
Finally, we comment on the very recent DMRG studies of Konik et al.~\cite{konik} 
on coupled TFIM systems in an effort at studying two dimensional coupled arrays 
of one dimensional systems. By starting with a reliable spectrum truncation 
procedure for a single TFIM chain (which relies on the underlying continuum 
1D theory being either conformally invariant or gapped but integrable), the 
authors then implement an improvement of their DMRG algorithm using first-order 
perturbative RG arguments. Their results for a $J_{\perp}$ coupling of the 
TFIM chains confirms the accuracy of the RPA analysis of Ref.(\cite{carr2}) 
and the present work in computing quantities like the single chain excitation 
gap (which is found to vanish at a critical $J_{\perp}$) and dispersion of 
excitations in the coupled system as a function of $J_{\perp}$. Their results 
indicate that the RPA method and the DMRG analysis agree very well upto values 
of $J_{\perp}$ of the order of 
the gap. This method also appears to give accurate values of critical 
exponents related to the ordering transition. Thus, this numerical approach 
appears to provide a confirmation of the interplay of the QCP in 
the TFIM and the transverse coupling in driving the dimensional crossover and 
deconfinement transition. Such an approach, therefore, holds much promise for 
the numerical investigations of such phenomena in systems with similar 
ingredients.

\section{Conclusions}
To conclude, we have studied a model of strongly correlated coupled 
quasi-1D systems at $1/4$-filling using an effective 
pseudospin TFIM model.  Using a bosonisation and RG analysis for a 2-leg 
ladder model, we find two different types
of charge/spin ordered ground states at $1/4$-filling. Transverse 
bond-fermion hopping is found to stabilise a new, gapped (insulating) 
phase characterised by interchain two-particle coherence of a type 
resembling orbital antiferromagnetism~\cite{kusmartsev,tsvelik}. 
The spin fluctuations are described by a $S=1/2$ Heisenberg  
ladder-type model for all cases studied here: the spin excitations are 
always massive. Away from this filling, either intra- or interchain 
superconductivity in a gapped spin background is found to be the stable 
ground state. We also find the existence of an intermediate gapless 
phase lying in between two gapped, charge-ordered phases (characterised 
by the relevant couplings $\tV$ and $\tp$ respectively) in the RG phase 
diagram of our model. 
\par
We followed this up with a RPA analysis of the case when many such effective TFIM 
systems are coupled. We find that a transverse bond-fermion (i.e., two particle) 
hopping coupling 
causes the ordered phase to extend to finite temperatures, with the phase 
boundary ending at a $T=0$ QCP. Interestingly, the critical quantum fluctuations 
at the QCP, together with the transverse coupling $t_{\perp}$, are found to drive a 
deconfinement transition (at $T=0$) together with a dimensional crossover 
characterised by strong dispersion of the excitations of any single TFIM into 
the transverse directions. The gapless higher dimensional system in the 
quantum critical region lying just above the QCP is in conformity with our finding 
of a critical phase in the 2-leg ladder model driven by $t_{\perp}$ discussed 
above. Similar conclusions are 
also reached for the effects of the transverse coupling $J_{\perp}$ on 
effective TFIM chains which are critical. The 
mechanism driving the dimensional crossover and deconfinement appear to be 
generic and lead us to believe that similar mechanism should exist for the 
case of other quantum critical systems in lower dimension when coupled to one 
another. Our present analyses are, however, especially relevant to coupled 
chain systems like the TMTSF organic systems as well as coupled ladder 
systems like $Sr_{14-x}Ca_{x}Cu_{24}O_{41}$ and $\alpha-NaV_{2}O_{5}$ or 
$\beta-Na_{0.33}V_{2}O_{5}$ (a superconductor) which exhibits charge/spin 
long range order at $x=0$ and superconductivity beyond 
under pressure and/or doping~\cite{tokura,yamauchi}.
    
\begin{acknowledgments}
SL and MSL thank the ASICTP and MPIPKS respectively for financial support. 
SL thanks F. Franchini, L. Dall'Asta, S. Basu, S. T. Carr, M. Fabrizio and 
A. Nersesyan for many discussions. 
\end{acknowledgments}

\end{document}